\input{aipcheck}

\documentclass[
    ,final            
  ]
  {aipproc}

\layoutstyle{8x11double}

\begin{document}

\title{A Search for PNe in Nearby Galaxies with SDSS Imaging Data}


\author{Alexei Y.\ Kniazev}{
  address={European Southern Observatory,
Karl-Schwarzschild-Strasse 2, 85748 Garching, Germany}
}

\author{Eva K.\ Grebel}{
  address={Astronomical Institute of the University of Basel,
      Venusstrasse 7, CH-4102 Binningen, Switzerland}
}

\author{Daniel B.\ Zucker}{
  address={Max-Planck-Institut f\"{u}r Astronomie, K\"{o}nigstuhl 17,
      D-69117 Heidelberg, Germany}
}

\author{Eric F.\ Bell}{
  address={Max-Planck-Institut f\"{u}r Astronomie, K\"{o}nigstuhl 17,
      D-69117 Heidelberg, Germany}
}

\author{Hans-Walter Rix}{
  address={Max-Planck-Institut f\"{u}r Astronomie, K\"{o}nigstuhl 17,
      D-69117 Heidelberg, Germany}
}

\author{David Mart\'{i}nez-Delgado}{
  address={Instituto de Astrofisica de Andaluc\'ia (CSIC),
Camino Bajo de Huetor, 24 18008 Granada, Spain}
}

\author{Hugh C.\ Harris}{
  address={US Naval Observatory, Flagstaff Station, P.O. Box 1149,
Flagstaff, AZ 860002-1149, USA}
}

\begin{abstract}
We present the latest results from our project to search for
new planetary nebulae in nearby galaxies using Sloan Digital
Sky Survey (SDSS) imaging data. Our method is based on photometric
criteria and can be applied to galaxies where PNe appear as
point sources. We applied these criteria to the whole area of M31
as scanned by SDSS, detecting 130 new PN candidates and 30 known
PNe. All selected PNe candidates are located in the outer regions
of M31. For 85 candidates follow-up spectroscopy was obtained
with the 2.2m telescope at Calar Alto Observatory. The observations
show that our method has a detection efficiency of about 82\%.
We discuss the 2D velocity field of the outer part of M31 based on our
observed PN data.  The PNe suggest an exponential disk scale length of
13 kpc along the minor axis. We discovered two PNe along the line of
sight to Andromeda NE, a very low surface
brightness giant stellar structure in the outer halo of M31.
These two PNe are located at projected distances of $\sim$48 kpc
and $\sim$41 kpc from the center of M31 and are the most distant PNe
in M31 found up to now.
Our data support the idea that Andromeda NE
is located at the distance of M31.
No PNe associated with other M31 satellites observed by the SDSS were found.
Applying our method to other SDSS regions
we checked data for the Local Group galaxies Sextans, Draco, Leo~I, Pegasus,
Sextans~B and Leo~A and recovered a known PN in Leo~A.
We re-measured its O/H
and for the first time determined abundances of N/H, S/H, He/H as well as
the electron number density $N_e$. We argue that the
PN progenitor was formed $\approx$1.5 Gyr ago
during the strongest episode of star formation in Leo\,A.
\end{abstract}

\maketitle


\section{Introduction}

Planetary nebulae (PNe) arise from low-mass stars, making them
excellent probes of the dynamics of low- to intermediate-mass
stars in nearby galaxies. Their emission lines can provide accurate
line-of-sight velocities within a minimum of telescope time. Therefore
spectroscopy of PNe can be used as a powerful tool for the study of
kinematics of nearby galaxies (e.g., Hurley-Keller et al.\,2004),
for the detection of new satellites (Morrison et al.\,2003)
and for the mapping of stellar accretion
streams around large galaxies (Merrett et al.\,2003).
The spectra of individual PNe provide chemical abundances of certain
elements, complementing
photometric or  spectroscopic metallicity information derived from old
red giants, young supergiants, or H\,{\sc ii} regions.
Searches for PNe are usually based on narrow-band imaging in the
H$\alpha$ and
[O\,{\sc iii}] $\lambda$5007 lines, in which PNe can emit 15--20\% of the
luminosity of the central star.

The Sloan Digital Sky Survey (SDSS) (York et al.\,2000) collects
imaging data in drift-scan mode in five bandpasses
($u, \ g, \ r, \ i$, and $z$; Fukugita et al.\,1996; Gunn et al.\,1998; Hogg et al.\,2001).
These data are then pipeline-processed
to measure photometric and astrometric properties
(Lupton et al.\,2002; Stoughton et al.\,2002; Smith et al.\,2002; Pier et al.\,2003).
Since the detected flux from PNe comes almost entirely from nebular emission
lines in the optical, the range of colors characteristic of the
PNe is defined by the ratios of these emission lines and their corresponding
contributions in different SDSS passbands. Some of these colors should
be similar to the colors of emission-line galaxies (ELGs)
and can be used for PN detection on the base
of SDSS photometry.

\section{The method}

We have developed a method to detect PN candidates in
SDSS imaging data based on photometric
criteria.
Using an SDSS scan of M31 reduced with the
standard pipeline (see Zucker et al\,2004)
and PNe from Nolthenius \& Ford\,(1987) and Jacoby \& Ford\,(1986)
we constructed a test sample of previously known PNe with SDSS parameters.
These were used to develop PN selection criteria
based on their SDSS colors and magnitudes (Kniazev et al.\,2004a).
We then applied these criteria to the whole area of M31
as scanned by the SDSS.
The star--galaxy separation for the SDSS is better than 90\% at
$r = 21.6^m$ (Abazajian et al.\,2003), but worsens for fainter magnitudes.
Thus we also applied the same selection criteria to extended objects.
All candidates were visually verified to minimize false detections such as
diffraction spikes of bright stars on SDSS images
and clearly extended objects.
Finally, we selected 105 PNe candidates that we labeled
``first priority'' (highly likely PNe) and 57 new PNe candidates labeled
``second priority'' (potential PNe; see Figure 1).
All PNe from the test sample were detected as candidates of the
first priority.
With our final criteria we then applied our method
to the whole SDSS data available on April 2004 and recovered a
PN in the Local Group galaxy Leo~A.  No PNe were detected in other
Local Group dwarfs covered by the SDSS: Sextans, Draco, Leo~I, Pegasus and
Sextans~B.
But for Leo~I, Pegasus and Sextans~B only photometry
for the outer parts is available from the SDSS pipeline due to crowding.

\begin{figure}
  \resizebox{0.45\textwidth}{!}{\includegraphics[angle=-90]{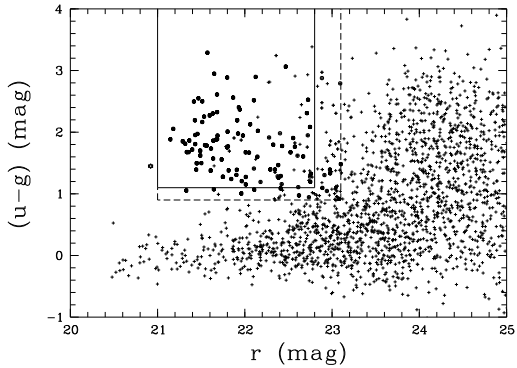}}
  \caption{%
Color-magnitude diagram for stellar sources from the SDSS M31 data.
All selected PN candidates of first priority
are located within the region delineated by the solid lines.
The dashed lines
show the softer criteria for the selection of second priority candidates.
All PNe from the test sample of previously known PNe
as well as true PNe confirmed with spectroscopic
follow-up observations are shown with filled circles.
All PN candidates with follow-up observations that did not reveal
obvious emission lines are marked by crossed circles.
The one PN in Leo~A is indicated by a star symbol.
   }
\end{figure}

We estimated possible contamination by point sources such as QSOs and stars,
analyzing their distribution as given by Richards et al.\,(2002),
and potential contamination by background emission-line galaxies,
using data from Kniazev et al.\,(2004b).
We find that our criteria select star-like
sources far from both the QSO and stellar loci. There are
only a few points located in our areas of interest,
which would yield perhaps 1--10 objects per 1000 deg$^2$.
The numbers of ELGs in our sample should also be extremely
small, since PNe stand out as being bluer in $(g-r)$ vs.\ $(r-i)$
and redder in  $(u-g)$ vs.\ $(g-r)$
due to the contribution of the very strong emission lines
H$\beta$ and [O\,{\sc iii}] $\lambda\lambda$4959,5007 in the SDSS
$g$-band and
H$\alpha$ in the SDSS $r$-band.

\begin{figure}
  \resizebox{0.45\textwidth}{!}{
   \includegraphics[angle=-90]{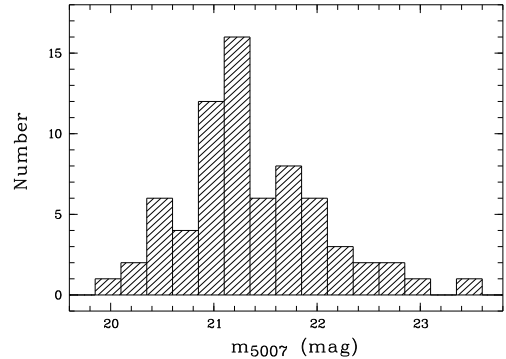}
}
  \caption{\label{fig:LF}
M31's PN luminosity function from this work.
   }
\end{figure}

\begin{figure}
  \resizebox{0.45\textwidth}{!}{
   \includegraphics[angle=-90]{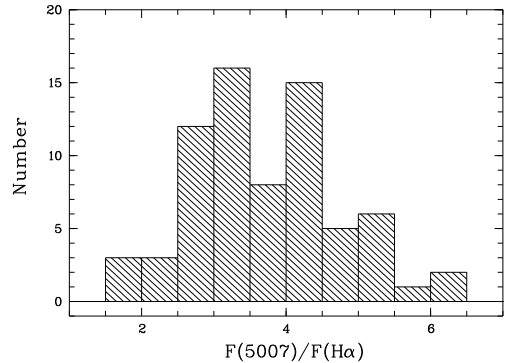}
}
  \caption{\label{fig:lines_ratio}
Observed [O~{\sc iii}] $\lambda$5007 to H$\alpha$ line ratio.
   }
\end{figure}

\begin{figure}
  \resizebox{1.\textwidth}{!}{\includegraphics[]{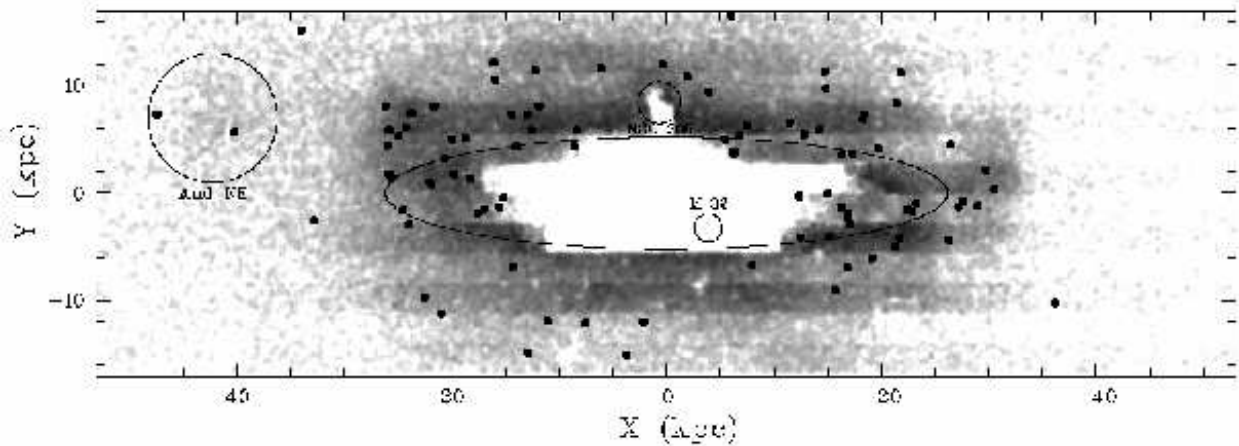}}
  \caption{\label{fig:Obs_final}
The spatial density of SDSS-detected stars is shown as a greyscale plot
in the coordinates of the SDSS scans.
The standard SDSS pipeline does not work properly in overcrowded fields,
resulting in the lack of data in the central area of M31.
The spatial distribution of newly-discovered PNe from the M31 SDSS data
is overplotted by filled circles.
The black ellipse indicates the approximate extent and inclination of
the bright M31 disk.
The large open circles mark the locations of NGC\,205, M32 and And NE.
The scale is plotted assuming a distance of 760 kpc to M31.
}
\end{figure}

\section{Spectral follow-up observations}

Spectroscopic follow-up
observations of a subset of the detected M31 PNe candidates
and of the PN in Leo~A were
carried out with the 2.2\,m telescope at Calar Alto Observatory
in October and May 2004.
We used the Calar Alto Faint Object Spectrograph (CAFOS).
In 8 nights of observations under variable
weather conditions
a total of 85 PN candidates were observed.
During these runs a long slit of variable width
depending on the seeing
and a G-100 grism were used.
The effective wavelength coverage was $\lambda$\,4200 --
$\lambda$\,6800\,\AA\ with
dispersion $\sim$1.9--2\,\AA/pix and
a spectral resolution $\sim$4--6\,\AA\ (FWHM).
All data were reduced using MIDAS and IRAF and all emission lines were
measured with the method described in detail in Kniazev et al.\,(2000)
and Kniazev et al.\,(2004b).

\section{Results}

\subsection{New PNe in M31}

Out of the observed 85 PN candidates from the M31 SDSS data
70 turned out to be genuine PNe, which implies a
total detection efficiency of 82\%.
Color-magnitude diagrams of the selected sources from SDSS M31 data
and the results of our follow-up observations are shown in Figure~1.
Figure~\ref{fig:LF} shows our PN luminosity function
in terms of m$_{5007}$ magnitudes calculated with the
standard equation from Jacoby\,(1989).
With our method we have found PNe over a range of 3 mag in the
m$_{5007}$-band,
the same magnitude range as
in the earlier searches with narrow-band filters (Hurley-Keller et al.\,2004).

The spatial distribution of the newly discovered PNe is shown
in Figure~\ref{fig:Obs_final}.  In part, they coincide with various
well-known
morphological features like
{\it the Northern spur, the NE Shelf, the NGC\,205 Loop,
the G1 clump, etc.}\, which provides an opportunity to study
abundances and velocities of these substructures of the outer part of M31.
We discovered two PNe with projected
locations near the center of And NE (Zucker et al\,2004).
This small number of PNe is consistent with the number to be expected
in a stellar structure of this luminosity ($\sim 5\times10^6 L_\odot$).
With their projected distances of $\sim$48 kpc and $\sim$41 kpc from the
center
of M31 these are the most distant PNe possibly belonging to M31 found up
to now.

\begin{figure}
  \resizebox{0.68\textwidth}{!}{\includegraphics[angle=-90]{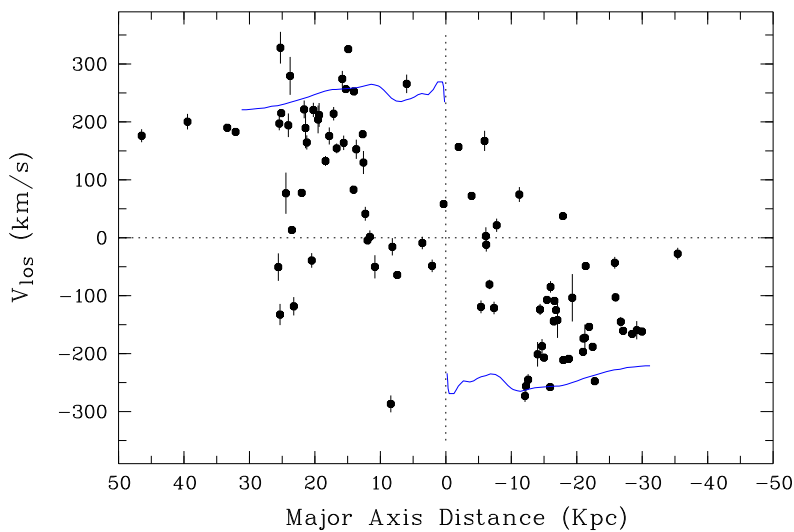}}
  \caption{\label{fig:gen_vels}
Velocity-distance diagram for newly discovered PNe in M31.
The X-axis represents the projected distance along
M31's major axis; the Y-axis shows the line-of-sight velocities.
Filled black circles are new PNe from our survey.
Errors of velocity determination are shown with bars.
The line is the H\,{\sc i} rotation curve from Kent\,(1989).
}
\end{figure}

The velocity distribution for all newly-discovered PNe along the major
axis of M31 is shown in Figure~\ref{fig:gen_vels}.
It is worth noting that most of them are rotationally supported and
trace the disk and/or bulge of M31, but do not show the signature of
a kinematically hot halo
where all objects would have random velocities.
The calculated density profile of the new PNe shows an exponential decline
with a scale length of $\alpha = 12.6$~kpc
along the minor axis of M31, which is very close to $\alpha = 13.7$~kpc
found by Irwin et al.\,(2005)
using the Isaac Newton Telescope Wide Field Camera survey of M31.

The velocities of the two PNe in Andromeda NE are close to each other
and agree well with the velocities of stars in this area (Ibata et al.\,2005).
They also fit in well with
the continuation of the H\,{\sc i} rotation curve from Kent\,(1989)
(see Figure~\ref{fig:gen_vels}).
Altogether with the results of our analysis of the spectroscopic line ratios
our PN data support the idea that Andromeda NE
is located at the distance of M31
and has [Fe/H] $\approx -0.7$.

\subsection{The PN in Leo~A}

Leo~A is a nearby dIrr at a distance of 800 kpc (Dolphin et al.\,2002) and
a member of the Local Group.
Strobel et al.\,(1991) detected an unresolved emission-line object as a PN candidate,
which was then observed spectroscopically by
Skillman et al.\,(1989), confirming its nature
as a PN. These authors found an oxygen abundance of
12+log(O/H) = 7.30$\pm$0.16 for this PN.  This was the only
spectroscopic measurement of the metallicity of Leo\,A to date.
Our value of 12+log(O/H) = 7.38$\pm$0.14
is in a good agreement with the earlier measurement.
Translating this into [Fe/H] yields
$\approx$ $-$1.28$\pm$0.14 ($\sim$5.2\% solar), significantly more
metal-rich than the photometrically derived metallicity of Leo\,A's old
population ($-$2.1 dex, Grebel et al.\,2003).
We used our data to also measure for the first time the abundances of
12+log(N/H) = 6.70, 12+log(S/H) = 4.60 and 12+log(He/H) = 11.05
as well as the electron number density $N_e = 1800$ for this PN.
Following the method described in Kniazev et al.\,(2005)
we calculated a mass of (M/M$_{\odot} \le$1.5) and an age of
(t$_{MS} <$1.6 Gyr) for the progenitor of the PN.
It would then have formed at the time when Leo\,A's star formation rate
showed a marked increase (Tolstoy\,1996).


\begin{theacknowledgments}
The Sloan Digital Sky Survey (SDSS) is a joint project of The University of
Chicago, Fermilab, the
Institute for Advanced Study, the Japan Participation Group, The Johns
Hopkins University, the
Max-Planck-Institute for Astronomy (MPIA), the Max-Planck-Institute for
Astrophysics (MPA),
New Mexico State University, Princeton University, the United States Naval
Observatory, and the
University of Washington. Apache Point Observatory, site of the SDSS
telescopes, is operated by the
Astrophysical Research Consortium (ARC).

Funding for the project has been provided by the Alfred P. Sloan Foundation,
the SDSS member
institutions, the National Aeronautics and Space Administration, the
National
Science Foundation, the
U.S. Department of Energy, the Japanese Monbukagakusho, and the Max Planck
Society. The SDSS Web site is http://www.sdss.org/.
\end{theacknowledgments}



\bibliographystyle{aipprocl} 




\end{document}